\documentstyle[12pt]{article}
\begin{document}

\title{Predictive Models of Large Neutrino Mixing Angles
\footnote{Supported 
by Department of Energy}}

\author{{\bf S.M.Barr}\\
Bartol Research Institute\\ University of Delaware\\
Newark, DE 19716}

\date{BA-96-24}
\maketitle

\begin{abstract}

Several experimental results could be interpreted as evidence
that certain neutrino mixing angles are large, of order unity.
However, in the context of grand unified models the neutrino
angles come out characteristically to be small, like the KM angles.
It is shown how to construct simple grand-unified models
in which neutrino angles are not only large but completely
predicted with some precision. Six models are presented for
illustration.

\baselineskip = .30in

\end{abstract}

\newpage

\baselineskip = .30in

\section{Introduction}

There are hints that some neutrino mixing angles may be
large. One interpretation of atmospheric neutrino data$^1$
suggests that the mixing angle between $\nu_{\mu}$ and
$\nu_{\tau}$ is of order unity.$^2$ There is also a large-angle
solution$^3$ to the MSW explanation$^4$ of the solar neutrino 
problem.$^5$ However, in unified theories of quark and lepton masses
there is a tendency for the leptonic mixing angles, like the
quark mixing angles, to come out small. In particular,
they tend to come out proportional to powers (either $1$ or
$\frac{1}{2}$) of the small intergenerational mass ratios.

In a recent paper$^6$ a general idea was proposed which gives
in a simple and natural way small KM angles and large
neutrino mixing angles in the context of unified theories.
This general idea has the additional virtue of explaining
why the hierarchy among the up quarks is larger than that among 
the down quarks and leptons. In that same paper$^6$ it was shown
that this idea could be combined with the idea of quark and lepton
mass-matrix ``textures" to give highly predictive schemes in
which the full $3 \times 3$ unitary mixing matrix of the neutrinos
is accurately predicted. In this paper we present a set of
five new models which (together with an example given in Ref. 6)
realize these ideas, and which give definite
and distinguishable predictions for the neutrino mixing angles.
These models not only illustrate the possibilities of this
approach, but demonstrate that at least within this framework
an experimental determination of the neutrino mixing angles
can settle the question of the origin of the pattern of light fermion
masses.

\vspace{0.5cm}

\section{The General Idea}

The general idea can be simply explained in the context of SU(5).
Consider a model where the fermions are in the representations
$(\overline{{\bf 5}}_i + {\bf 10}_i + {\bf 1}_i) + 
(\overline{{\bf 10}}'_i + {\bf 10}'_i)$, where $i = 1,2,3$ is a family index.
Let the fermion mass terms be

\begin{equation}
\begin{array}{ccl}
{\cal L}_{{\rm mass}} & = & \sum_{i,j} l^{(0)c}_i (L_0)_{ij} l^{(0)}_j
+ \sum_{i,j} d^{(0)c}_i (D_0)_{ij} d^{(0)}_j \\
& & \\ 
& & + \sum_{i,j} u^{(0)c}_i (U_0)_{ij} u^{(0)}_j +
\sum_{i,j} {\nu}^{(0)c}_i (N_0)_{ij} {\nu}^{0}_j \\
& & \\
& & + \sum_i M_i \overline{{\bf 10}}'_i {\bf 10}'_i + 
\sum_i m_i \overline{{\bf 10}}'_i {\bf 10}_i.
\end{array}
\end{equation}

\noindent
The fields $l^{(0)c}_i$, $u^{(0)}_i$, $d^{(0)}_i$, and
$u^{(0)c}_i$ belong to the ${\bf 10}$ of SU(5) denoted
${\bf 10}_i$. The fields $l^{(0)}_i$, $\nu^{(0)}_i$, and
$d^{(0)c}_i$ belong to the $\overline{{\bf 5}}$ denoted 
$\overline{{\bf 5}}_i$. In addition there is a set of vectorlike pairs
denoted $\overline{{\bf 10}}'_i + {\bf 10}'_i$. 

Note that we write the matrices, $L_0$ etc., so that
the left-handed fermions are to the right and the
left-handed antifermions are to the left. This will be
the convention throughout this paper. 
The matrices $L_0$, $D_0$, $U_0$, and $N_0$ do not have to satisfy
the minimal SU(5) relations, but will in general come from effective operators
that involve the GUT-scale breaking of SU(5). That is why we
write these mass terms using $SU(3) \times SU(2) \times
U(1)$ representations instead of $SU(5)$ multiplets. We imagine
in this paper that these four matrices are constrained by some
kind of family symmetry to have a ``texture" form. Moreover,
we assume that for each of these matrices all the non-zero elements 
are of the same order of magnitude. That is to say, the matrices
$L_0$, $D_0$, $U_0$, and $N_0$
do not exhibit a significant intergenerational hierarchy.

The intergenerational hierarchies come from the mixing with 
the $\overline{{\bf 10}}'_i + {\bf 10}'_i$ in the following
way. As is clear from Eq. (1), 
the $\overline{{\bf 10}}'_i$ gets a Dirac mass, assumed
to be superheavy, with the linear combination $(\cos \theta_i
{\bf 10}'_i + \sin \theta_i {\bf 10}_i) \equiv {\bf 10}_{{\rm heavy}, i}$,
where $\tan \theta_i = m_i/M_i$. The orthogonal combination
$(- \sin \theta_i {\bf 10}'_i + \cos \theta_i {\bf 10}_i) \equiv
{\bf 10}_{{\rm light},i}$ is light and contains the 
Weak-scale-mass observable states, $u_i$, $d_i$, $u^c_i$, and $l^c_i$.
Thus, the ${\bf 10}_i$, which contains the fields $l^{(0)c}_i$, $u^{(0)}_i$,
$d^{(0)}_i$, and $u^{(0)c}_i$ appearing in 
Eq.(1) is related to the true low-mass states by

\begin{equation}
{\bf 10}_i = \cos \theta_i {\bf 10}_{{\rm light},i} +
\sin \theta_i {\bf 10}_{{\rm heavy},i}.
\end{equation}

\noindent
That means that we can write the mass matrices of the 
light quarks and leptons as

\begin{equation}
\begin{array}{ccl}
L & = & H \; L_0 \\
D & = & \; \; \; \; D_0 \; H \\
U & = & H \; U_0 \; H \\
N & = & \; \; \; \; N_0, 
\end{array}
\end{equation}

\noindent
where

\begin{equation}
H = \left( \begin{array}{ccc}
\cos \theta_1 & & \\
& \cos \theta_2 & \\
& & \cos \theta_3
\end{array}
\right) \equiv \left( \begin{array}{ccc}
\epsilon_1 & & \\
& \epsilon_2 & \\
& & \epsilon_3
\end{array}
\right).
\end{equation}

\noindent
From these equations it is clear that if there is a hierarchy
$\epsilon_1 \ll \epsilon_2 \ll \epsilon_3$, that is to say
if $M_1/m_1 \ll M_2/m_2 \ll M_3/m_3$, then there will be
an intergenerational hierarchy both among the masses of the down quarks and
among those of the charged leptons that goes as $\epsilon_1 ~:~ \epsilon_2 ~:~
\epsilon_3$, and a hierarchy among the up quark masses
that goes as $\epsilon_1^2 ~:~ \epsilon_2^2 ~:~ \epsilon_3^2$.
From Eq.(3) one also sees that the mixing angles among the
left-handed quarks are of order the hierarchy factors,
that is $V^{KM}_{ij} \sim \epsilon_i/\epsilon_j$, $i<j$,
while the mixing angles among the left-handed leptons are
of order unity (since no factor of $H$ appears to the right of
$L_0$ and $N_0$ in Eq. (3)).

To be able to actually predict the neutrino mixing angles
from the knowledge we already possess about the quark and lepton
masses and the KM angles there must be a symmetry that relates
$N_0$ to $L_0$, $D_0$, and $U_0$. This would suggest enlarging
the symmetry group to SO(10). In that case the representations
$\overline{{\bf 5}}_i + {\bf 10}_i + {\bf 1}_i$ are unified
into ${\bf 16}_i$, and the $\overline{{\bf 10}}'_i + {\bf 10}'_i$
could be unified either into ${\bf 45}_i$ or into  
$\overline{{\bf 16}}'_i
+ {\bf 16}'_i$.

Unifying the $\overline{{\bf 10}}'_i + {\bf 10}'_i$ into
${\bf 45}_i$ would mean that only the ${\bf 10}_i$, and not the
$\overline{{\bf 5}}_i$, mixed with
the vectorlike states, since the ${\bf 45}$ contains 
$\overline{{\bf 10}} + {\bf 10}$ but not $\overline{{\bf 5}}
+ {\bf 5}$. This would lead to the structure shown in Eqs. (1)-(3),
and is the situation that is assumed in this paper.

However, the other possibility is also interesting. If the
$\overline{{\bf 10}}'_i + {\bf 10}'_i$ are contained in
$\overline{{\bf 16}}'_i + {\bf 16}'_i$, then there
are also ${\bf 5}'_i + \overline{{\bf 5}}'_i$ with which the
$\overline{{ 5}}_i$ (that contain the 
$l^{(0)}_i$, $\nu^{(0)}_i$, and $d^{(0)c}_i$) mix. If, instead
of just mass terms like $M_i \overline{{\bf 16}}'_i {\bf 16}'_i
+ m_i \overline{{\bf 16}}'_i {\bf 16}_i$ analogous to the
terms in Eq. (1), one had as well terms where $M_i$ and
$m_i$ were replaced by the VEV of an adjoint Higgs field, ${\bf 45}_H$,
which pointed along the SU(5)-singlet generator of SO(10), then
different mixing matrices, $H_{10}$, $H_{\overline{5}}$, and $H_1$,
would exist for the ${\bf 10}$, $\overline{{\bf 5}}$, and ${\bf 1}$
of SU(5). Then one would have $L = H_{10} L_0 H_{\overline{5}}$, 
$D = H_{\overline{5}} D_0 H_{10}$, $U = H_{10} U_0 H_{10}$,
and $N = H_1 N_0 H_{\overline{5}}$. We shall not explore this
possibility in this paper.

Assuming that the $\overline{{\bf 10}}'_i + {\bf 10}'_i$
are unified into ${\bf 45}_i$, the 
terms involving the vectorlike fermions in Eq.(1) become
at the SO(10) level $(\sum_i M_i {\bf 45}_i {\bf 45}_i +
\sum_i m_i {\bf 45}_i {\bf 16}_i \langle \overline{{\bf 16}}_H
\rangle )$. These terms will also generate GUT-scale right-handed
neutrino masses, since the ${\bf 45}_i$ contains singlets,
${\bf 1}'_i$, which will mix with the singlets ${\bf 1}_i$.
It is easy to see by integrating out the superheavy
singlets ${\bf 1}'_i$ and ${\bf 1}_i$ that the Majorana
mass matrix of the light, left-handed neutrinos takes the form

\begin{equation}
(M_{\nu})_{ij} = \frac{4}{5} (N_0^T)_{ik} m^{-1}_k
M_k m^{-1}_k (N_0)_{kj},
\end{equation}

\noindent
or

\begin{equation}
M_{\nu} = \frac{4}{5} N_0^T \tilde{H} M^{-1} \tilde{H} N_0,
\end{equation}

\noindent
where $\tilde{H} \equiv {\rm diag}(\cot \theta_1,
\cot \theta_2, \cot \theta_3) \simeq H = {\rm diag}(\epsilon_1,
\epsilon_2, \epsilon_3)$ and the $\frac{4}{5}$ is an SO(10) Clebsch. 
$\tilde{H}$ has a hierarchy similar to that of $H$.

From the forms in Eqs. (3) and (4) it is straightforward to derive explicit
expressions for the mass ratios of the charged quarks and leptons
and for the KM angles; and from Eq. (6) one can in the same way
derive expressions for the neutrino mixing angles as we shall see.

From Eqs. (3) and (4) it is apparent that the elements of $U$ have a hierarchy
$U_{ij} \propto \epsilon_i \epsilon_j$. That is to say, there is
a hierarchy in both the rows and columns of $U$. Therefore,

\begin{equation}
m_c/m_t \cong \left( \epsilon_2/\epsilon_3 \right)^2 
\frac{\det_{23} U_0}{(U_0)_{33}^2},
\end{equation}

\begin{equation}
m_u/m_t \cong \left( \epsilon_1/\epsilon_3 \right)^2 
\frac{\det U_0}{(U_0)_{33} \det_{23} U_0}.
\end{equation}

\noindent
The down-quark matrix, $D = D_0 H$, has a hierarchy
among its columns, but not among its rows.
Thus it is convenient to define the column vectors
$(\vec{D}_j)_i = (D_0)_{ij}$. Then it is straightforward
to show that

\begin{equation}
m_s/m_b \cong (\epsilon_2/\epsilon_3) \frac{\left| \vec{D}_2 \times
\vec{D}_3 \right|}{\left| \vec{D}_3 \right|^2},
\end{equation}

\begin{equation}
m_d/m_b \cong (\epsilon_1/\epsilon_3) \frac{\left| 
\vec{D}_1 \cdot \vec{D}_2 \times \vec{D}_3 \right| \left| \vec{D}_3
\right|}
{\left| \vec{D}_2 \times \vec{D}_3 \right|^2}.
\end{equation}

\noindent
Since both $U$ and $D$ have hierarchies among their
{\it columns}, the rotations among the {\it left-handed}
$u_i$ and $d_i$ required to diagonalize the mass matrices
will be small, proportional to hierarchy factors
$\epsilon_i/\epsilon_j$. One can write down the leading
order (in $\epsilon_i/\epsilon_j$, $i < j$) expressions
for the Kobayashi-Maskawa angles in a simple form.

\begin{equation}
\begin{array}{ccl}
V_{cb} & \cong & \left( \frac{\epsilon_2}{\epsilon_3} \right)
\left[ \frac{\vec{D}_2 \cdot \vec{D}_3}{(\vec{D}_3)^2}
- \frac{U_{0,32}}{U_{0,33}} \right], \\
& & \\
V_{us} & \cong &  \left( \frac{\epsilon_1}{\epsilon_2} \right)
\left[ \frac{\vec{D}_1 \cdot \vec{D}_2 - \vec{D}_1 \cdot
\hat{D}_3 \vec{D}_2 \cdot \hat{D}_3}{|\vec{D}_2 \times
\hat{D}_3 |^2} - \frac{U_{0,33} U_{0,21} - 
U_{0,31} U_{0,23}}{\det_{23} U_0} \right], \\
& & \\
V_{ub} & \cong & \left( \frac{\epsilon_1}{\epsilon_3} \right)
\left[ \frac{\vec{D}_1 \cdot \vec{D}_3}{(\vec{D}_3)^2}
- \frac{U_{0,31}}{U_{0,33}} +\left(
\frac{U_{0,33} U_{0,21} - 
U_{0,31} U_{0,23}}{\det_{23} U_0} \right) \left( 
\frac{\vec{D}_2 \cdot \vec{D}_3}{(\vec{D}_3)^2}
- \frac{U_{0,32}}{U_{0,33}} \right) \right].
\end{array}
\end{equation}

\noindent
Note that $V_{ub} \sim V_{us} V_{cb}$.

The expressions for the mass ratios of the charged
leptons are similar in form to those of the down quarks,
except that $L = H L_0$ has a hierarchy among its {\it rows}
and not its columns. Thus it is convenient to define
the row vectors $(\vec{L}_i)_j \equiv (L_0)_{ij}$.
In terms of these

\begin{equation}
m_{\mu}/m_{\tau} \cong (\epsilon_2/\epsilon_3) \frac{\left| \vec{L}_2 \times
\vec{L}_3 \right|}{\left| \vec{L}_3 \right|^2},
\end{equation}

\begin{equation}
m_e/m_{\tau} \cong (\epsilon_1/\epsilon_3) \frac{\left| 
\vec{L}_1 \cdot \vec{L}_2 \times \vec{L}_3 \right| \left| \vec{L}_3
\right|}
{\left| \vec{L}_2 \times \vec{L}_3 \right|^2}.
\end{equation}

In discussing the neutrino mixing angles let us assume for 
the moment that the masses $M_i$ in Eq. (1) 
are all of the same order, so that
the hierarchy among the $\epsilon_i \equiv \cos \theta_i 
= M_i/\sqrt{m_i^2 + M_i^2}$ is due to a hierarchy among the
$m_i$. Then it is apparent from Eqs. (5) and (6) that one has
effectively as a neutrino Dirac mass matrix
$N_{\nu, {\rm eff}} \equiv \tilde{H} N_0$. This, like
$L = H L_0$ has a hierarchy among its rows but not among
its columns. Therefore, the leptonic analogue of the KM
matrix has mixing angles of order unity, and to leading
order the small parameters $\epsilon_i/\epsilon_j$, $i<j$,
do not enter. It is straightforward to show that

\begin{equation}
V_{{\rm lepton}} = V_N^{\dag} V_L,
\end{equation}

\noindent
where

\begin{equation}
V_L \cong \left( \frac{\vec{L}_2 \times \hat{L}_3}{| \vec{L}_2
\times \hat{L}_3 |}, \frac{\vec{L}_2 - \vec{L}_2 \cdot \hat{L}_3
\hat{L}_3}{| \vec{L}_2 \times \hat{L}_3 |}, \hat{L}_3 \right),
\end{equation}

\noindent
and
 
\begin{equation}
V_N \cong \left( \frac{\vec{N}_2 \times \hat{N}_3}{| \vec{N}_2
\times \hat{N}_3 |}, \frac{\vec{N}_2 - \vec{N}_2 \cdot \hat{N}_3
\hat{N}_3}{| \vec{N}_2 \times \hat{N}_3 |}, \hat{N}_3 \right).
\end{equation}
 
If, indeed, the $M_i$ are all of the same order, with the hierarchy
being among the $m_i$, then the rows of $N_{\nu, {\rm eff}} =
\tilde{H} N_0$ have a hierarchy of order $\epsilon_1 ~:~ 
\epsilon_2 ~:~ \epsilon_3$. In that case the corrections to
Eq. (14) are easily shown to be $\delta (V_{{\rm lepton}})_{ij} \sim
(\epsilon_i/\epsilon_j)^2$, $i<j$. On the other hand,
it could be that the $m_i$ are all of the same order, with
the hierarchy being among the $M_i$. In that case one can write Eq. (4)
more usefully as $M_{\nu} = \frac{4}{5} N_0^T \tilde{H}^{\frac{1}{2}}
m^{-1} \tilde{H}^{\frac{1}{2}} N_0$. From this it is evident
that the corrections to the expression for $V_{{\rm lepton}}$ given
in Eq. (14) are of order $\delta (V_{{\rm lepton}})_{ij} \sim
\epsilon_i/\epsilon_j$. In either case, the corrections,
as we shall see, are small enough in realistic cases to mean that
the predictions of particular models are sufficiently sharp to allow
them to be distinguished.

\section{Texture Models}

\noindent
{\bf (a) An example: Model Aa}

We will construct models in which the matrices $L_0$, $D_0$,
$U_0$, and $N_0$ have a common ``texture" form.
An example, which was given in Ref. 6, is the following:

\begin{equation}
L_0 = \left[ \begin{array}{ccc} 
& -3D & \\ D & -C & \\ & B/2 & A 
\end{array} \right],
\end{equation}

\begin{equation}
D_0 = \left[ \begin{array}{ccc} 
& D & \\ -3D & C/3 & \\ & B/2 & A 
\end{array} \right],
\end{equation}

\begin{equation}
U_0 = \left[ \begin{array}{ccc} 
& D & \\ D & C/3 & \\ & -B/2 & A 
\end{array} \right] \tan \beta,
\end{equation}

\begin{equation}
N_0 = \left[ \begin{array}{ccc} 
& -3D & \\ 5D & -C & \\ & -B/2 & A 
\end{array} \right] \tan \beta,
\end{equation}

\noindent
with $B/A = 0.4$, $C/A = 0.75$, $D/A = 0.06$, 
$\epsilon_2/\epsilon_3 = 0.08$, and $\epsilon_1/\epsilon_3
= 0.02$. This gives the following fit to the quark
and lepton masses and mixings:
$m_{\tau}/m_b = 1.02$, $m_{\mu}/m_s \cong 3.0$, 
$m_e/m_d \cong 0.33$, $m_{\mu}/m_{\tau} \cong 0.06$,
$m_e/m_{\mu} \cong 5 \times 10^{-3}$, $m_c/m_t
\cong 1.6 \times 10^{-3}$, $m_u/m_c \cong 3.5 \times 10^{-3}$,
$V_{us} \cong 0.22$, $V_{ub} \cong 0.002$, and $V_{cb}
\cong 0.03$. (These quantities are all defined at
the unification scale.)

With these values of the parameters, one finds, using Eqs. (14)-(16)
that 

\begin{equation}
V_{{\rm lepton}}^{(Aa)} = \left(
\begin{array}{ccc}
0.95 & 0.3 & -0.088 \\
-0.3 & 0.87 & -0.39 \\
0.032 & 0.4 & 0.92
\end{array}
\right).
\end{equation}

\noindent
The superscript $(Aa)$ is the name we give to this
particular model in this paper, for reasons that will become apparent
later. It should be noted that all four matrices in Eqs. (17)-(20)
have the same form, which can be written

\begin{equation}
F_0 \propto  \left(
\begin{array}{ccc}
0 & D \; X[f] & 0 \\
D \; X[f^c] & C \; (B-L)[f^c] & B \; I_{3R}[f] \\
0 & B \; I_{3R}[f^c] & A
\end{array}
\right),
\end{equation}

\noindent
with $F = L$, $D$, $U$, or $N$. The quantities $B-L$, $I_{3R}$, 
and $X$ are just generators of SO(10). $B-L$ and
$I_{3R}$ (the third generator of $SU(2)_R$) are 
conventionally normalized. $X$ is the SU(5)-singlet
generator that is normalized so that the ${\bf 10}$, 
$\overline{{\bf 5}}$, and ${\bf 1}$ contained in the
${\bf 16}$ have the charges $1$, $-3$, and $5$ respectively.
Sometimes we shall use generators, $\overline{Q}$, 
consistently normalized so that tr$|_{16} \overline{Q}^2 = 1$.
Then $\overline{B-L} = \frac{\sqrt{3}}{4}(B-L)$, 
$\overline{I_{3R}} = \frac{1}{\sqrt{2}}I_{3R}$, and $\overline{X} = 
\frac{1}{4\sqrt{5}} X$.
Writing Eq. (22) in terms of those normalized generators

\begin{equation}
F_0 = \left( \begin{array}{ccc}
0 & \overline{D} \; \overline{X}[f] & 0 \\
\overline{D} \; \overline{X}[f^c] & \overline{C} \;
(\overline{B-L})[f^c] & \overline{B} \; \overline{I_{3R}}[f] \\
0 & \overline{B} \; \overline{I_{3R}}[f^c] & \overline{A}
\end{array}
\right).
\end{equation}

\noindent
One has then that $\overline{B}/\overline{A} = \sqrt{2} B/A = 0.566$,
$\overline{C}/\overline{A} = 1.73$, and $\overline{D}/\overline{A}
= 0.537$. Since we are attempting to explain the
intergenerational hierarchies by the mixing parameters
$\epsilon_2/\epsilon_3$ and $\epsilon_1/\epsilon_3$,
it is most natural if the ratios of these barred parameters
are of order unity, as indeed we see that they are in this model.
One can regard this as an encouraging success of this approach.

Another success of this approach is the fact that the same form
can be used in all four matrices, $L_0$, $D_0$, $U_0$, and $N_0$.
In usual texture models, using the same form for $U$ and $D$
either gives $V_{cb} \cong 0$ (if $U_{32}$ and $D_{32} = 0$), or
$V_{cb} \sim \sqrt{m_s/m_b}$ (if $U_{32}$ and $D_{32} \neq 0$),
which is much too large. Here, even with the same form for $D_0$ and $U_0$
(up to the group theory factors),
$V_{cb} \sim \epsilon_2/\epsilon_3 \sim m_s/m_b$, which is of the
correct order.

The generators of SO(10) can be introduced into the
form $F_0$ simply through higher-dimension effective
operators obtained from integrating out vectorlike fermion
representations. Consider the following set of terms

\begin{equation}
{\cal L}' = a \overline{{\bf 16}} \Omega_{\tilde{Q}} {\bf 16}
+ b \overline{{\bf 16}} \Omega_Q {\bf 16}_i + c {\bf 16}
{\bf 16}_j {\bf 10}_H.
\end{equation}

\noindent
here $i$ and $j$ are not dummy indices but are particular
values of the indices. $\Omega_Q$ is either an adjoint
(${\bf 45}$) of Higgs fields, whose VEV is proportional to
the SO(10) generator $Q$, or it is an explicit mass
or singlet Higgs, in which case $Q$ is just the identity.
The same possibilities exist for $\Omega_{\tilde{Q}}$.
Both $\Omega_Q$ and $\Omega_{\tilde{Q}}$ are taken to be of
the GUT scale. It is easy to see that if one integrates out
the superheavy fermion $\overline{{\bf 16}}$ and its
superheavy partner ${\bf 16}_{{\rm heavy}} 
\propto a \langle \Omega_{\tilde{Q}} \rangle {\bf 16} + b \langle
\Omega_Q \rangle {\bf 16}_i$ one obtains the following
effective operator

\begin{equation}
{\cal O} = c \frac{b \langle \Omega_{Q({\bf 16}_i)} \rangle}
{\sqrt{| b \langle \Omega_{Q({\bf 16}_i)} \rangle |^2
+ | a \langle \Omega_{\tilde{Q}({\bf 16}_i)} \rangle |^2}}
{\bf 16}_i {\bf 16}_j {\bf 10}_H.
\end{equation}

\noindent
Here $Q({\bf 16}_i)$ is the value of $Q$ acting on the
appropriate component of the ${\bf 16}_i$. Let us assume
that $|b \langle \Omega_Q \rangle |^2 \ll | a \langle
\Omega_{\tilde{Q}} \rangle |^2$. Then the operator
is approximately

\begin{equation}
{\cal O} \propto \frac{Q({\bf 16}_i)}{\tilde{Q}({\bf 16}_i)}
{\bf 16}_i {\bf 16}_j {\bf 10}_H.
\end{equation}

\noindent
Consider the contributions of this operator to the matrix
$F_0$. There are two contributions.

\begin{equation}
{\cal O}_f = \frac{Q(f^c)}{\tilde{Q}(f^c)} f^c_i f_j v^{(f)}
+ \frac{Q(f)}{\tilde{Q}(f)} f_i f^c_j v^{(f)}.
\end{equation}

\noindent
If $i \neq j$ one has the combination of generators 
$Q(f^c)/\tilde{Q}(f^c)$ appearing in the $ij$ element
of $F_0$, and $Q(f)/\tilde{Q}(f)$ appearing in the
$ji$ element. For example, in the model described
above (the (Aa) model, cf. Eq. (23)) one can get the 23 and 32 elements
of the right form by taking $i=3$, $j=2$, $Q= I_{3R}$,
$\tilde{Q} = 1$; and one can get the 12 and 21 elements
by taking $i=2$, $j=1$, $Q = X$, and $\tilde{Q} = 1$.

For $i=j$, the
operator in Eq. (26) leads to the combination of
generators $[Q(f^c)/\tilde{Q}(f^c) + Q(f)/\tilde{Q}(f)]$
appearing in the $ii$ element of $F_0$. In the example
model, one gets the 33 element in a trivial way by taking
$i=j=3$ and $Q= \tilde{Q}= 1$. The 22 element of that model
requires more discussion.  
For $i=j=2$ take $Q = B-L$ and $\tilde{Q}$ to be a linear
combination of $I_{3R}$ and $1$. (That is, $\Omega_{\tilde{Q}}$
is a linear combination of an adjoint with VEV proportional to
$I_{3R}$ and an explicit mass.) In particular, take

\begin{equation}
\frac{Q[f^c]}{\tilde{Q}[f^c]} + \frac{Q[f]}{\tilde{Q}[f]} =
\frac{(B-L)[f^c]}{\eta^{-1} I_{3R}[f^c] + 1} 
+ \frac{(B-L)[f]}{\eta^{-1} I_{3R}[f] + 1},
\end{equation}

\noindent
with $\eta \ll 1$. Since the left-handed fermions have
$I_{3R} = 0$, the second term is just $(B-L)[f]$.
But $I_{3R}[f^c] = \pm \frac{1}{2}$, so the first term
is $\pm 2 \eta (B-L)[f^c] \ll 1$. For small $\eta$ we can
therefore neglect the first term and the combination
of generators is approximately $(B-L)[f] = - (B-L)[f^c]$.

\vspace{0.5cm}

\noindent
{\bf (b) A Second Example: Model Bb} 

Our second example is given by

\begin{equation}
L_0 = \left[ \begin{array}{ccc} 
E/2 &  & \\ D/2 & C/2 & \\ & B/2 & A 
\end{array} \right],
\end{equation}

\begin{equation}
D_0 = \left[ \begin{array}{ccc} 
-3E/2 & & \\ D/2 & -C/6 & \\ & -3B/2 & A 
\end{array} \right],
\end{equation}

\begin{equation}
U_0 = \left[ \begin{array}{ccc} 
3E/2 & & \\ -D/2 & -C/2 & \\ & 3B/2 & A 
\end{array} \right] \tan \beta,
\end{equation}

\begin{equation}
N_0 = \left[ \begin{array}{ccc} 
-E/2 & & \\ -D/2 & -C/10 & \\ & -B/2 & A 
\end{array} \right] \tan \beta,
\end{equation}

\noindent
with $B/A = 0.36$, $C/A = 4.6$, $D/A =1.15$, $E/A = 0.10$,
$\epsilon_2/\epsilon_3 = 0.027$, $\epsilon_1/\epsilon_3 = 0.0062$.
This gives the same values for the mass ratios and KM angles
as model (Aa), except that the value of $V_{ub}/(V_{us}V_{cb})$
comes out to be about 0.25 instead of 0.3. The four mass matrices
have the common form

\begin{equation}
F_0 = \left( \begin{array}{ccc}
E \left( \frac{I_{3R}}{B-L} \right)[f^c] & D \; I_{3R}[f] & \\
D \; I_{3R}[f^c] & C \; \left( \frac{I_{3R}}{X} \right)[f^c] &
B \; \left( \frac{I_{3R}}{B-L} \right)[f] \\
& B \; \left( \frac{I_{3R}}{B-L} \right)[f^c] & A 
\end{array} \right).
\end{equation}

\noindent
If we use the consistently normalized SO(10) generators, then
we find $\overline{B}/\overline{A} = 0.22$, $\overline{C}/
\overline{A} = 0.727$, $\overline{D}/\overline{A} = 1.63$,
$\overline{E}/\overline{A} = 0.061$. Except for the last
quantity all the parameter ratios are of order unity.

The neutrino-mixing matrix obtained from 
Eqs. (14) - (16) is 

\begin{equation}
V_{{\rm lepton}}^{(Bb)} = \left(
\begin{array}{ccc}
-0.78 & 0.56 & 0.27 \\
-0.62 & -0.76 & -0.215 \\
0.086 & -0.34 & 0.94 
\end{array} \right).
\end{equation}

In both the models discussed so far $U_{0,31} = 0 = \vec{D}_1
\cdot \vec{D}_3$. In this case Eqs. (11) simplify
to give 

\begin{equation}
\frac{V_{ub}}{V_{us} V_{cb}} \cong 
\left[ \left( \frac{\vec{D}_1 \cdot \vec{D}_2}{|\vec{D}_2 \cdot
\hat{D}_3 |^2} \right)/ \left( \frac{U_{0,33} U_{0,21}}{\det_{23} U_0}
\right) - 1 \right]^{-1}.
\end{equation}

\vspace{0.5cm}

\noindent
{\bf (c) Six Models and Predictions for Neutrino Mixing}

All the models presented here have the following general
form

\begin{equation}
F_0 = \left( \begin{array}{ccc}
c_{11}^F E & c_{12}^F D & 0 \\
c_{21}^F D & c_{22}^F C & c_{23}^F B \\
0 & c_{32}^F B & c_{33}^F A 
\end{array} \right),
\end{equation}

\noindent
where, for each $ij$, $i \neq j$, there is a pair of
generators $Q$, $\tilde{Q}$, such that 
$c_{ij}^F = Q[f^c]/\tilde{Q}[f^c]$, 
$c_{ji}^F = Q[f]/\tilde{Q}[f]$, and where
for each $ii$ there is a pair of generators
$Q$, $\tilde{Q}$ such that $c_{ii}^F = (Q[f^c]/\tilde{Q}[f^c]
+ Q[f]/\tilde{Q}[f])$. In Table I we present for each
of the six models the group-theoretical factors
that appear in each of the entries of $F_0$.
In Table II are given the numerical values of the parameters
of the models that give good fits to the observed quark and
lepton masses and KM angles. In Table III are given
the neutrino-mixing matrices that are predicted
in each model from Eqs. (14)-(16), as well as the values of
$V_{ub}/(V_{us}V_{cb})$ that are predicted from Eqs. (11).

The reason for the names we have given the models
can be seen from Table I. The models with the same 
capital letter have the same form in the 2-3 block.
This means that they get the Georgi-Jarlskog ratio
$m_{\mu}/m_s \cong 3$ in the same way. There are three
such forms used, which we call $A$, $B$, and $C$.
Similarly, models with the same lower-case letter have the
same form in the 
1-2 block (more precisely, in the 11, 12, and 21 elements).
There are three such forms used which we have called $a$, $b$, and
$c$. These forms are arranged to give the other Georgi-Jarlskog
factor $m_e/m_d \cong \frac{1}{3}$.

From Table II we see that the values of the ratios $\overline{B}/
\overline{A}$, $\overline{C}/\overline{A}$, and $\overline{D}/
\overline{A}$ are of order unity, as is natural in this framework 
where the intergenerational hierarchies come from the
ratios $\epsilon_i/\epsilon_j$. In particular, the 18 entries
in Table II that give these three ratios for the six models are
all between $\frac{1}{10}$ and $10$. Indeed, 12 of these 18 numbers
are between $\frac{1}{2}$ and $2$. The values of $\overline{E}/
\overline{A}$ are somewhat smaller for the five models which have
this parameter, ranging from $0.058$ to $0.331$, and being typically about
$\frac{1}{10}$.

It should be noted that the signs of the entries in $V_{{\rm lepton}}$
shown in Table III are not individually of absolute significance.
First of all, a change in the sign of $\epsilon_i/\epsilon_j$
gives a change in the signs of certain fermion mass ratios and
KM angles. Since these are not known, one can get equally good
fits to the known data by assuming lepton mass ratios of various signs.
Thus, one can change the sign of any of the left or right-handed
lepton mass eigenstates and have essentially the same fit.
Therefore in $V_{{\rm lepton}}$ the sign of any row or column
can be changed and still correspond to a model which fits the
known data.

\section*{References}

\begin{enumerate}

\item Kamiokande Collaboration: K.S. Hirata {\it et al.},
Phys. Lett.{\bf B205}, 416 (1988); Phys. Lett. {\bf B280},
146 (1992); Y. Fukuda {\it et al.}, Phys. Lett. {\bf B335},
237 (1994); IMB Collaboration: D. Caspar {\it et al.}, 
Phys. Rev. Lett. {\bf 66}, 2561 (1991); R. Becker-Szendy
{\it et al.}, Phys. Rev. {\bf D46}, 3720 (1992).

\item For a review of the theory, see T.K. Gaisser, in TAUP 93,
Nucl. Phys. (Proc. Suppl.), {\bf B35}, 209 (1994).

\item For a recent update, see e.g., N. Hata, S. Bludman,
and P. Langacker, Phys. Rev. {\bf D49}, 3622 (1994); P.I. Krastev
and Yu. Smirnov, Phys.Lett. {\bf B338}, 282 (1994).

\item S.P. Mikheyev and A.Yu. Smirnov, Sov. J. Nucl. Phys.
{\bf 42}, 913 (1986); L. Wolfenstein, Phys. Rev. {\bf D17},
2369 (1978).

\item J. Bahcall, {\it Neutrino Astrophysics}, Cambridge University
Press, Cambridge, England (1989).

\item K.S. Babu and S.M. Barr, BA-95-56, hep-ph/9511446

\end{enumerate}

\newpage

\noindent
{\large\bf Table I:} The combinations of generators
appearing in each entry of the fermion mass matrix 
$F_0$ for each of the six models. These generators act on
the left-handed $f^c$ for entry $ij$ ($i \neq j$),
on the left-handed $f$ for entry $ji$, and on the
left-handed $f$ for diagonal entries $ii$. For the 22 column
the entries ``$(B-L)_f$" stand for the expression
$(B-L)/(\eta^{-1} I_{3R} + 1)$ where $\eta$ is a small
parameter. In model Bc, the fourth column
entry (distinguished by a (**)) corresponds to $ij = 12$ and
not $ij = 21$ as for the other models.
\vspace{0.2in}

$$\begin{array}{c|c|c|c|c|c}\hline
& & & & & \\
{\rm Entry} \; ij \rightarrow & 33 & 32 & 22 & 21 & 11 \\
\hrulefill & & & & & \\
\downarrow {\rm Model} & & & & & \\
\hline
Aa & 1 & I_{3R}/1 & (B-L)_f & X/1 & - \\
\hline
Ab & 1 & I_{3R}/1 & (B-L)_f & I_{3R}/1 & I_{3R}/(B-L) \\
\hline
Ac & 1 & I_{3R}/1 & (B-L)_f & X/1 & I_{3R}/(B-L) \\
\hline
Bb & 1 & I_{3R}/(B-L) & I_{3R}/X & I_{3R}/1 & I_{3R}/(B-L) \\
\hline
Bc & 1 & I_{3R}/(B-L) & I_{3R}/X & 1/X \; (**) & I_{3R}/(B-L) \\
\hline
Cb & 1/X & (B-L)/1 & 1 & I_{3R}/1 & I_{3R}/(B-L) \\
\hline
\end{array}
$$

\newpage

\noindent
{\large\bf Table II:} The values of the parameters that
give a good fit to the known quark and lepton masses and 
the KM angles for each of the six models. (The barred
quantities are the coefficients of the SO(10)-consistently
normalized generators.)

\vspace{0.2in}

$$
\begin{array}{c|c|c|c|c|c|c}
\hline
& & & & & & \\
{\rm Model} & B/A & C/A & D/A & E/A & \epsilon_2/\epsilon_3 & 
\epsilon_1/\epsilon_3 \\
& (\overline{B}/\overline{A}) & (\overline{C}/\overline{A}) & 
(\overline{D}/\overline{A}) & (\overline{E}/\overline{A}) & & \\
\hline
Aa & 0.4 & 0.75 & 0.06 & - & 0.08 & 0.02 \\
& (0.566) & (1.73) & (0.537) & & & \\
\hline
Ab & 0.4 & 0.75 & -0.66 & 0.094 & 0.08 & 0.0064 \\
& (0.566) & (1.73) & (-0.933) & (0.058) & & \\
\hline
Ac & 0.4 & 0.75 & 0.3 & 0.54 & 0.08 & 0.0035 \\
& (0.566) & (1.73) & (2.68) & (0.331) & & \\
\hline
Bb & 0.36 & 4.6 & 1.15 & 0.10 & 0.027 & 0.0062 \\
& (0.22) & (0.727) & (1.63) & (0.061) & & \\
\hline 
Bc & 0.36 & 4.6 & -1.31 & -0.344 & 0.027 & 0.005 \\
& (0.22) & (0.727) & (-0.147) & (-0.205) & & \\
\hline 
Cb & -0.3 & 0.033 & 0.147 & 0.025 & 0.28 & 0.0157 \\
& (-6.2) & (0.30) & (1.85) & (0.139) & & \\
\hline
\end{array}
$$

\newpage

\noindent
{\large\bf Table III:} The predicted values of the
leptonic mixing angles and of $V_{ub}$ in each of the
six models.

\vspace{0.2in}

$$ 
\begin{array}{c|c|c}
\hline 
& & \\
{\rm Model} & V_{{\rm lepton}} & V_{ub}/(\sin \theta_c V_{cb}) \\
& & \\
\hline
Aa & \left( \begin{array}{ccc} 
0.95 & 0.3 & -0.088 \\
-0.3 & 0.87 & -0.39 \\
0.032 & 0.4 & 0.92 
\end{array}
\right) & 0.3 \\
\hline
Ab & \left( \begin{array}{ccc}  
0.68 & 0.72 & -0.157 \\
-0.72 & 0.60 & -0.35 \\
-0.157 & 0.35 & 0.92
\end{array}
\right) & 0.5 \\
\hline
Ac & \left( \begin{array}{ccc}
0.72 & 0.60 & -0.345 \\
-0.68 & 0.72 & -0.17 \\
0.145 & 0.355 & 0.92 
\end{array}
\right) & 0.25 \\
\hline
Bb & \left( \begin{array}{ccc}
-0.78 & 0.56 & 0.27 \\
-0.62 & -0.76 & -0.215 \\
0.086 & -0.34 & 0.94 
\end{array}
\right) & 0.25 \\
\hline
Bc & \left( \begin{array}{ccc}
-0.472 & -0.840 & -0.268 \\
0.878 & -0.418 & -0.229 \\
0.081 & 0.344 & 0.936 
\end{array}
\right) & 0.56 \\
\hline
Cb & \left( \begin{array}{ccc}
0.80 & 0.58 & 0.163 \\
-0.53 & 0.54 & 0.645 \\
0.29 & -0.604 & 0.75 
\end{array}
\right) & 0.56 \\
\hline
\end{array}
$$

\end{document}